\documentclass[aps,twocolumn,groupedaddress,showpacs]{revtex4}
\usepackage{graphicx}
\begin{document}
% You should use BibTeX and revtex.bst for references
\bibliographystyle{apsrev}

%\preprint{}

\title{Kondo Lattice Behavior in the Ordered Dilute Magnetic Semiconductor Yb$_{14-x}$La$_x$MnSb$_{11}$}
% Optional argument for running titles on pages
%\title[]{}


% \affiliation command. The \affiliation command MUST follow the
\author{B.C. Sales$^1$}
\author{P. Khalifah$^{1,2}$}
\author{T. P. Enck$^1$}
\author{E. J. Nagler$^1$}
\author{R. E. Sykora$^3$}
\author{R. Jin$^1$}
\author{D. Mandrus$^1$}

\affiliation{$^1$Condensed Matter Sciences Division, Oak Ridge
National Laboratory, Oak Ridge, TN 37831}
\affiliation{$^2$Chemistry Department, University of
Massachusetts, Amherst, MA 01003}
\affiliation{$^3$Chemical
Sciences Division, Oak Ridge National Laboratory, Oak Ridge, TN
37831}

\date{\today}

\begin{abstract}
We report Hall, magnetic, heat capacity and doping studies from
single crystals of Yb$_{14}$MnSb$_{11}$ and
Yb$_{13.3}$La$_{0.7}$MnSb$_{11}$. These heavily doped
semiconducting compounds are ferromagnetic below 53 K and 39 K
respectively. The renormalization of the carrier mass from 2m$_e$
near room temperature to 15m$_e$ at 5 K , plus the magnetic
evidence for partial screening of the Mn magnetic moments suggest
that these compounds represent rare examples of an underscreened
Kondo lattice with T$_K \approx$ 350 K.
\end{abstract}
\pacs{75.30.Mb, 72.15.Qm, 75.50.Cc}
%\maketitle must follow title, authors, abstract and PACS
\maketitle

Ferromagnetic semiconductors are envisioned as a key component of
many proposed spintronic devices that functionalize both the
electron charge and spin.\cite{1,2} There have been many theories
of the origin of ferromagnetism in doped semiconductors,\cite{3}
but it has been difficult to obtain a clean comparison between
theory and experiment because of problems associated with
clustering or phase separation of magnetic dopants in the
semiconducting host. In the present article we show that it is
productive to investigate the ferromagnetism in stoichiometric
compounds such as Yb$_{14}$MnSb$_{11}$ (T$_c$=53K). In this
heavily doped semiconducting compound the magnetic Mn atoms are at
a unique crystallographic site in the structure resulting in a
minimum Mn-Mn separation of 10 \AA. Using a combination of Hall,
heat capacity, magnetic data, and doping studies we are able to
show that the ferromagnetism is likely mediated by relatively
heavy quasiparticles that form due to the Kondo interaction. This
leads to the non-intuitive and surprising result that using
carrier tuning to increase T$_c$ can result in a reduced
saturation moment. Since the local environment of each Mn atom is
similar to that found in the heavily studied III-V semiconductors
\cite{4} (such as GaAs:Mn), the present results may also have
significant implications for III-V materials as well.

The compound Yb$_{14}$MnSb$_{11}$ was first synthesized by Chan
$\it {et}$ $\it {al.}$ in 1998.\cite{5} It is isostructural with a
large family of related Zintl compounds such as
Ca$_{14}$AlSb$_{11}$ and Ca$_{14}$MnSb$_{11}$\cite{6}
crystallizing with a tetragonal lattice in the space group $\it
{I41/acd}$. The structure of Yb$_{14}$MnSb$_{11}$ is shown in
Fig.\ 1, where for clarity only the Mn and four nearest neighbor
Sb atoms are shown. Previous magnetization and X-ray Magnetic
Circular Dichroism (XMCD) measurements indicate that Yb is
divalent in this compound with no magnetic moment.\cite{5,7}
Initial magnetization and magnetic susceptibility measurements
suggested a Mn$^{+3}$ ($\it {d}$$^4$) configuration.\cite{5,8} The
MnSb$_4$ tetrahedron is somewhat distorted with angles of
105.6$^\circ$ and 117.5$^\circ$ which were previously interpreted
as a Jahn-Teller distortion associated with a d$^4$
configuration.\cite{5} However, more recent XMCD measurements are
more consistent with a Mn$^{+2}$ (d$^5$) configuration with the
moment of one spin compensated by the antialligned spin of an Sb
5p hole from the Sb$_4$ cage surrounding the Mn.\cite{7} There is
also a substantial distortion of the AlSb$_4$ tetrahedra in the
isostructural compound Ca$_{14}$AlSb$_{11}$ implying that steric
effects cause distortions even in the absence of Jahn-Teller
effects.\cite{9} A d$^5$+hole (d$^5$+h) picture is also expected
from detailed electronic structure calculations on related
Ca$_{14}$MnBi$_{11}$ compound.\cite{10} This is in accord with the
understanding of the most heavily studied dilute magnetic
semiconductor (DMS) GaAs:Mn which is suggested by multiple
experiments to have also a d$^5$+h  Mn configuration.\cite{2,3,11}

We grew relatively large (up to 0.5 g) single crystals of
Yb$_{14}$MnSb$_{11}$ with and without various dopants (La, Te, Y,
Sc etc.) capable of modifying the carrier concentration using the
conditions worked out by Fisher $\it {et}$ $\it {al.}$\cite{8} for
pure Yb$_{14}$MnSb$_{11}$. The partial substitution of La for Yb
had the largest effect on T$_c$ and the hole carrier concentration
and hence of the doped samples we investigated these crystals in
the most detail. A single crystal structure refinement of a La
doped crystal (Bruker SMART APEX CCD X-ray diffractometer)
indicates a composition of Yb$_{13.3}$La$_{0.7}$MnSb$_{11}$ with
all of the La atoms substituting on an Yb site (a = 16.6613(6)
\AA, c = 21.9894(7) \AA). This is expected since La$^{+3}$ and
Yb$^{+2}$ ions are about the same size and are chemically similar.
Hall effect, resistivity, and heat capacity data were obtained
using a Physical Property Measurement System (PPMS) from Quantum
Design on oriented and thinned single crystals. Hall and
resistivity data were obtained using a standard six lead method
and either rotating the sample by 180$^\circ$ in a fixed magnetic
field or by sweeping the direction of the field from positive to
negative values. The Hall data were qualitatively the same
regardless of the orientation of the crystal with respect to the
current or field directions. Magnetization data were obtained
using a commercial SQUID magnetometer from Quantum Design.

 Magnetic data for single crystals of Yb$_{14}$MnSb$_{11}$ and
 Yb$_{13.3}$La$_{0.7}$MnSb$_{11}$ are shown in Fig.\ 2.  The
 partial substitution of La for Yb lowers T$_c$ by about 14 K, but
 results in an increase of the saturation moment from $\sim$
 4.2$\mu_B$ to 4.5$\mu_B$ per Mn.

\begin{figure}
\includegraphics[keepaspectratio=true, totalheight =3.0 in, width =
2.5 in]{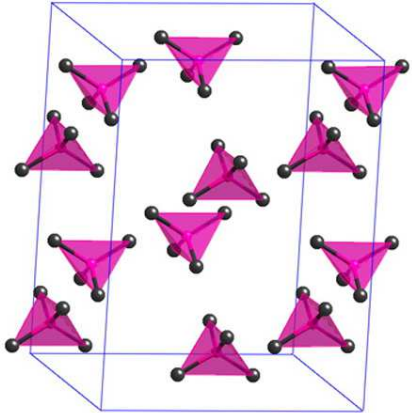}
 \caption{Structure of the tetragonal compound Yb$_{14}$MnSb$_{11}$ with $\it {a}$ = 16.61 \AA
and $\it {c}$ = 21.95 \AA. For clarity only the Mn atoms (red) and
the nearest neighbor Sb atoms (black) are shown. The closest Mn-Mn
distance is 9.98 \AA. }
\end{figure}

Heat capacity data from both materials are shown in Fig.\ 3. The
transitions to the ferromagnetic phases are evident at T$_c$ = 53
K (Yb$_{14}$MnSb$_{11}$ crystal) and T$_c$=39 K
(Yb$_{13.3}$La$_{0.7}$MnSb$_{11}$ crystal). The small
concentration of magnetic Mn atoms, and the relatively large
lattice contribution to the heat capacity data near T$_c$ make it
difficult to accurately determine the magnetic contribution to the
heat capacity data below T$_c$. The electronic contribution to the
heat capacity, $\gamma$T, can be reliably estimated, however by
using the heat capacity data from the lowest temperatures (T $\le$
5 K) and plotting C/T vs. T$^2$. This analysis yields $\gamma$
$\sim$ 58 mJ/K$^2$-mole Mn for the La doped crystal and $\gamma$
$\sim$ 120 mJ/K$^2$-mole Mn for the undoped crystal. The value of
$\gamma$ for the undoped crystal is very similar to the values
previously measured by Fisher $\it {et}$ $\it {al.}$\cite{8} and
Burch $\it {et}$ $\it {al.}$\cite{12} However, to determine the
significance of $\gamma$ requires a measure of the carrier
concentration and an estimate of the expected band mass.

\begin{figure}
\includegraphics[keepaspectratio=true, totalheight =3.0 in, width =
2.5 in]{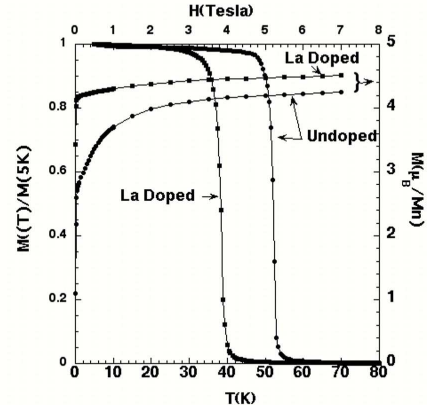} \caption{Normalized magnetization (left scale)
vs. temperature for crystals cooled in a field of 100 Oe. The
Curie temperature of the Yb$_{14}$MnSb$_{11}$ crystal is T$_c$ =
53 $\pm$ 1 K, a value consistent with values previously
reported.\cite{1,7} Magnetization (right scale) vs. applied
magnetic field for Yb$_{14}$MnSb$_{11}$ and
Yb$_{13.3}$La$_{0.7}$MnSb$_{11}$ single crystals. Although the
T$_c$ of the La doped crystal is lower, the saturation
magnetization is about 4.5$\mu_B$ per Mn as compared to about
4.2$\mu_B$ per Mn for the undoped crystals.}
\end{figure}

The carrier concentrations for the crystals are determined from a
detailed analysis of Hall data that takes into account the
anomalous Hall effect (AHE) associated with the ferromagnetism.
For many ferromagnets it has been experimentally
demonstrated\cite{13} that the Hall resistivity,  $\rho_{xy} =
R_0B + R_sM$, where R$_0$ is the "normal" Hall coefficient, B is
the magnetic field, M is the magnetization and R$_s$ is the
"anomalous" Hall coefficient. In the simplest situations R$_0$ is
proportional to 1/n, where n is the effective carrier
concentration. There are a variety of theories and models for the
AHE, which can be viewed as an additional current that develops in
the y direction in response to an electric field in the x
direction (magnetization in z direction).\cite{14,15,16} There can
be both intrinsic ($\rho_{xy} \propto \rho_{xx}^2$) and extrinsic
contributions ( $\rho_{xy} \propto \rho_{xx}$) to the AHE that are
usually separated based on how  $\rho_{xy}$ depends on the normal
resistivity, $\rho_{xx}$.\cite{13} Interest in spin currents for
spin based electronics, coupled with new theoretical insights into
the microscopic origin of R$_s$, have stimulated substantial
recent interest in the AHE. While the AHE of Yb$_{14}$MnSb$_{11}$
and related alloys are quite interesting, these data will be
discussed in more detail in a future publication. The measured
Hall resistivity for Yb$_{14}$MnSb$_{11}$ is shown in Fig.\ 4.
Analysis of these data (as described in the figure caption)
indicates that the carrier concentration is about 1.0 $\times$
10$^{21}$ to 1.35 $\times$ 10$^{21}$ holes/cm$^3$ above T$_c$,
depending on how the data are analyzed, but increasing to 1.9
$\times$ 10$^{21}$ holes/cm$^3$ for temperatures well below T$_c$.
All of the measured carrier concentrations are typical of a
heavily doped semiconductor and the values correspond to
approximately 1 hole/Mn (1 hole per Mn gives 1.3 $\times$
10$^{21}$ holes/cm$^3$). Doping the crystals with a small amount
of La ($\it {i.e.}$ Yb$_{13.3}$La$_{0.7}$MnSb$_{11}$) lowers the
measured hole concentration to about 4 $\times$ 10$^{20}$
holes/cm$^3$. This observation is consistent with the simple idea
that the extra electrons contributed to the compound when part of
the Yb$^{+2}$ ions are replaced by La$^{+3}$ ions fill some of the
holes in the Sb 5p bands. A crude estimate for the band gap of
$\sim$ 1 eV is obtained from the optical data of Burch $\it {et}$
$\it {al.}$\cite{12} on Yb$_{14}$MnSb$_{11}$, but it is not clear
if the carrier concentration in these materials can be reduced
enough to observe a gap with electrical transport measurements.

\begin{figure}
\includegraphics[keepaspectratio=true, totalheight =4.0 in, width =
3.0 in]{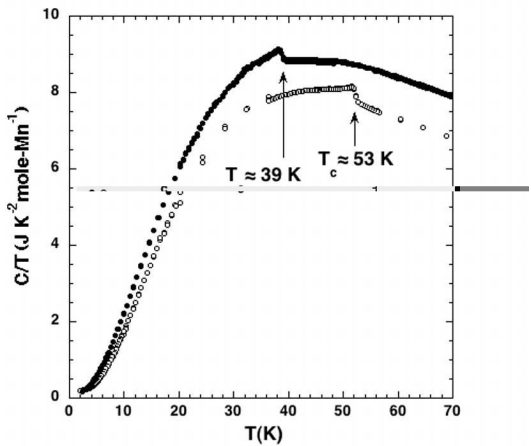} \caption{Heat capacity divided by temperature
(C/T) vs. temperature (T) for single crystals of
Yb$_{14}$MnSb$_{11}$ and Yb$_{13.3}$La$_{0.7}$MnSb$_{11}$. The
ferromagnetic phase transition for each crystal is noted. Analysis
of these data at low temperatures yield values for the electronic
heat coefficient of $\gamma$ = 120 $\pm$ 10 mJ/K$^2$-mole Mn, and
$\gamma$ = 58 $\pm$ 5 mJ/K$^2$-mole Mn for the undoped and
La-doped crystals, respectively. Heat capacity data on both
crystals were taken down to 0.5 K and in magnetic fields up to 14
T (not shown). The data were corrected for a small nuclear
Schottky contribution that was significant for temperatures below
1 K}
\end{figure}

The average Fermi wavevector, k$_F$, only depends on the number of
carriers per unit volume and is given by k$_F =(3\pi^2n)^{1/3}
\sim$ 0.36 $\AA^{-1}$. The Fermi energy, E$_F$, from the free
electron model is given by h$^2k_F^2/(4\pi^2m_e)$ which gives
E$_F$ = 0.5 eV. If the band mass, m$_b \approx 1.8m_e$, is used
(as measured from optical data\cite{12} or estimated from
electronic structure calculations,\cite{10} or estimated\cite{17}
from the room temperature Seebeck coefficient of +44 $\mu$V/K)
E$_F \sim$ 0.28 eV. The measured values for the carrier
concentration indicate a substantial mass enhancement for the
carriers of about 15m$_e$ for the undoped crystal and 11m$_e$ for
the La doped crystal. These mass enhancement values suggest a
Kondo temperature T$_K \sim$ 370 K. Similar mass enhancement
values and estimates of the Kondo temperature were obtained by
Burch $\it {et}$ $\it { al.}$\cite{12} for the undoped crystal
using optical and heat capacity data. In addition, Burch $\it
{et}$ $\it {al.}$\cite{12} were able to map the evolution of the
enhanced carrier mass as both the temperature and frequency were
lowered toward zero. They also noted that the optical data were
qualitatively similar to that found in many Kondo lattice
compounds.

\begin{figure}
\includegraphics[keepaspectratio=true, totalheight =3.5 in, width =
3.2 in]{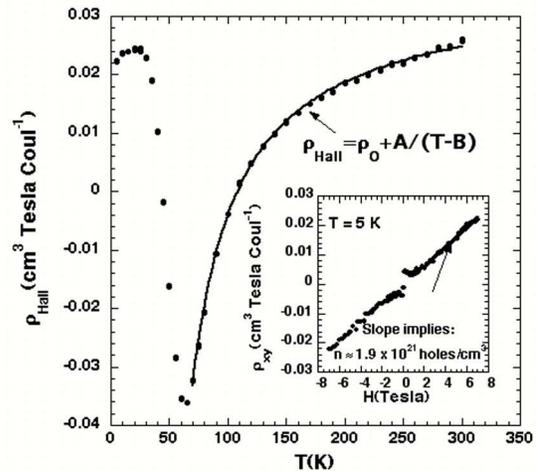} \caption{Hall resistivity ($\rho_{xy}$) vs.
temperature for Yb$_{14}$MnSb$_{11}$ measured in a field of 7
Tesla. The carrier concentration is estimated by fitting the data
above T$_c \sim$ 53 K to a constant plus a Curie-Weiss Law.
Similar data (not shown) were also taken with a field of 1 Tesla.
This analysis yields an average value for the carrier
concentration above T$_c$ of 1.35 $\times 10^{21}$ holes/cm$^3$.
Comparable values for the carrier concentration are obtained if
the measured magnetization data are used. If the Hall resistivity
above T$_c$ is analyzed using the expression derived for Kondo
lattice compounds above the coherence temperature\cite{22} (the
anomalous part of the Hall resistivity R$_s$M is replaced by
C$\rho_{xx}$M, where C is a constant) the data yields a
temperature independent (75 K $\le T \le$ 300K) carrier
concentration of 1.03 $\pm$ 0.05 $\times$ 10$^{21}$ holes/cm$^3$.
The carrier concentration well below T$_c$ is estimated by
plotting $\rho_{xy}$/B vs M/B using the measured magnetization
data as shown in the inset. This analysis yields a carrier
concentration of 1.9 $\times$ 10$^{21}$ holes/cm$^3$. It appears
that from just above T$_c$ (T $\approx$ 75 K) to well below T$_c$,
the carrier concentration jumps by about 0.9 $\times$ 10$^{21}$
holes cm$^{-3}$.}
\end{figure}

\begin{figure}
\includegraphics[keepaspectratio=true, totalheight =3.0 in, width =
3.0 in]{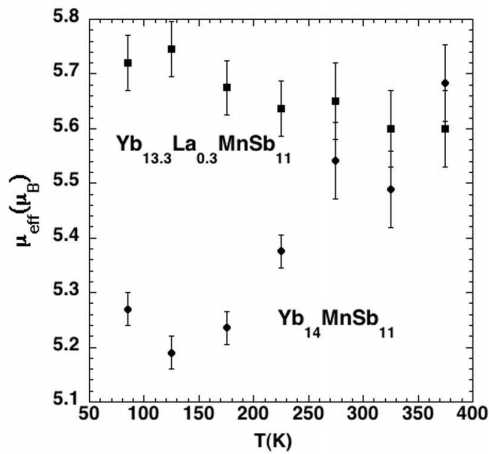} \caption{Effective magnetic moment versus
temperature extracted from DC susceptibility data taken in a field
of 1 T. The Curie-Weiss temperature, T$_{CW}$, was fixed for each
material and the data was fit to $\chi$(cm$^3$/mole Mn) = $\chi_0$
+ C/(T-T$_{CW})$ over 50 K intervals, with
$\mu_{eff}$=(8C)$^{1/2}$.}
\end{figure}

The basic Kondo interaction describes the antiferromagnetic
interaction between a localized spin (in this case a Mn d$^5$
configuration) and the mobile carriers, which for
Yb$_{14}$MnSb$_{11}$ would be 5p holes from the Sb bands.  A
material with a regular array of these  spins (or Kondo centers)
is referred to as a Kondo lattice compound. While the Kondo
interaction is often associated with the formation of a
non-magnetic singlet and a large enhancement of the carrier mass,
this same interaction is also responsible for RKKY (carrier
mediated exchange) coupling between spins which can result in
magnetic order.\cite{18} In rare cases, such as CePdSb, both
effects can occur which results in ferromagnetic order but with a
reduced saturation moment.\cite{19} Dietl $\it {et}$ $\it {
al.}$\cite{11} have shown that in Mn doped III-V semiconductors
there is a strong tendency for the Kondo interaction to bind holes
from neighboring group V $\it {p}$ bands to the $\it {d}$ shell of
the Mn$^{+2}$ impurity resulting in a composite object with a net
magnetic moment of 4$\mu_B$. The binding energy of the holes in
Yb$_{14}$MnSb$_{11}$ is relatively weak and is roughly given by
T$_K \approx$ 370 K. The standard estimate for the size of the
Kondo compensation cloud is hv$_F$/2$\pi$k$_B$T$_K$ $\approx$ 40
\AA, which also indicates weak binding. At temperatures near T$_K$
the composite quasiparticles should begin to break up with less
compensation of the Mn $\it {d}$$^5$ moments by the 5$\it {p}$
holes. A careful analysis of the magnetic susceptibility data from
Yb$_{14}$MnSb$_{11}$ (Fig.\ 5) indicates an increase of the
effective moment for temperatures near T$_K$. The experimental
data from the doped crystals are also consistent with this
picture. The replacement of Yb by La reduces the number of
carriers and weakens the indirect exchange coupling resulting in a
lower T$_c$. The lower number of holes also reduces the magnetic
screening of the Mn $\it {d}$$^5$ moments which results in a
larger saturation moment and a larger effective magnetic moment,
as is observed. Our work suggests that the unusual $\it {d}$$^5$+h
magnetism of Mn atoms may be intrinsic to many dilute magnetic
semiconductors. This leads to the non-intuitive and surprising
result that using carrier tuning to increase T$_c$ can result in a
reduced saturation moment. The present data also support the idea,
proposed by Burch $\it {et}$ $\it {al.}$,\cite{12} that
Yb$_{14}$MnSb$_{11}$ is an underscreened Kondo lattice compound. A
general characteristic of these compounds is that part of the
entropy associated with the local magnetic moment (Mn spins) is
removed by a transfer of part of the local $\it {d}$ character to
the conduction band.\cite{20,21} This transfer begins near the
coherence temperature, T$^*$, and results in a large
renormalization of the mass of quasiparticles at the Fermi energy
and modification of the Fermi surface below T$^*$. Burch $\it
{et}$ $\it {al.}$\cite{12} estimated that T$^*$ is near the
ferromagnetic transition temperature (T$^* \approx$ 50 K). This is
consistent with the present Hall data that indicates a jump in the
carrier concentration (which is a measure of the topology of the
Fermi surface) from about 1.2 $\times$ 10$^{21}$ holes cm$^{-3}$
above T$_c$ to 1.9 $\times$ 10$^{21}$ holes cm$^{-3}$ well below
T$_c$. Further studies of these materials are clearly needed,
however, to substantiate the Kondo lattice interpretation of the
present data. We also believe that further investigations of
natural DMS, such as Yb$_{14}$MnSb$_{11}$, will provide additional
insights into to the complex physics of carrier mediated magnetism
that will be essential to the understanding and design of actual
spintronic devices.

% If in twocolumn mode, this environment will move to single column
% format so that long equations can be displayed. Use
% Note: this may cause bad behavior if this occurs near a page
% break - this can be worked around by adding an explicit \pagebreak.
%\begin{widetext}
%\end{widetext}

% If you have acknowledgments this puts in the proper section
\begin{acknowledgments}
It is a pleasure to acknowledge enlightening discussions with
Kenneth Burch, David Singh, Victor Barzykin, Igor Zutic, Thomas
Schulthess and Steve Nagler. Oak Ridge National Laboratory is
managed by UT-Battelle LLC, for the U.S. Department of Energy
under contract DE-AC05-00OR22725.
\end{acknowledgments}

%references

% figures follow here or may be put into the text as floats.
% Use the graphics or
% graphicx packages distributed with LaTeX2e. See the LaTeX Graphics
% Companion by Michel Goosens, Sebastian Rahtz, and Frank Mittelbach
% for instance.
% \begin{figure}
% \label{}
% \includegraphics[]{}
% \caption{}
% \end{figure}

% tables follow here or maybe be put in the text
% \begin{table}
% \label{}
% \caption{}
% \end{tabular}
% \end{table}

\end{document}